\documentclass{article}
\usepackage{spconf,amsmath,graphicx}
\usepackage{amssymb}
\usepackage{multirow}
\usepackage{color}
\usepackage[english]{babel}
\usepackage[utf8]{inputenc}
\usepackage{balance}
\usepackage{algorithm}
\usepackage[noend]{algpseudocode}
\usepackage{amsmath}
\usepackage{breqn}
\usepackage[printwatermark]{xwatermark}

\usepackage{soul}
\usepackage{todonotes}

\title{Class-Conditional Defense GAN Against End-to-End Speech Attacks}
\name{Mohammad Esmaeilpour, Patrick Cardinal, Alessandro Lameiras Koerich}
\address{\'{E}cole de Technologie Sup\'{e}rieure (\'{E}TS),
D\'{e}partement de G\'{e}nie Logiciel et des TI\\
1100 Notre-Dame W, Montr\'{e}al, H3C 1K3, Qu\'{e}bec, Canada\\
{\small{\texttt{mohammad.esmaeilpour.1@ens.etsmtl.ca, \{patrick.cardinal, alessandro.koerich\}@etsmtl.ca}}}
}

\begin{document}

\ninept
\maketitle
\begin{abstract}
In this paper we propose a novel defense approach against end-to-end adversarial attacks developed to fool advanced speech-to-text systems such as DeepSpeech and Lingvo. Unlike conventional defense approaches, the proposed approach does not directly employ low-level transformations such as autoencoding a given input signal aiming at removing potential adversarial perturbation. Instead of that, we find an optimal input vector for a class conditional generative adversarial network through minimizing the relative chordal distance adjustment between a given test input and the generator network. Then, we reconstruct the 1D signal from the synthesized spectrogram and the original phase information derived from the given input signal. Hence, this reconstruction does not add any extra noise to the signal and according to our experimental results, our defense-GAN considerably outperforms conventional defense algorithms both in terms of word error rate and sentence level recognition accuracy.
\end{abstract}


\section{Introduction}
\label{sec:intro}
The threat of adversarial attacks has been well characterized in the domains of audio and speech recognition \cite{schonherr2018adversarial,esmaeilpour2020detection}. Classifiers either trained on raw signals or their corresponding 2D representations (i.e., spectrograms) are quite vulnerable against carefully crafted adversarial examples and this poses a serious concern about safety and reliability of these models \cite{yakura2018robust}. In a big picture, there are two main directions in studying adversarial attacks for speech signals: (i) generalizing strong attack algorithms developed for natural images in computer vision domain to spectrograms, taking advantage of their lower computational complexity \cite{koerich2020cross,esmaeilpour2019robust}; (ii) developing end-to-end attacks which require dealing directly with raw input signals \cite{carlini2018audio,qin2019imperceptible}. In this paper, we focus on the latter for defense purposes since it is closely related to a black-box attack scenario in real-life applications.

Although there are different implementations for end-to-end attacks, they unanimously use variants of the logarithmic distortion metric $l_{\text{ dB}_{\vec{x}}}(\delta)=l_{\text{dB}}(\delta)-l_{\text{dB}}(\vec{x})$ \cite{carlini2018audio}, which measures the loudness in dB of an adversarial example $\vec{x}_{adv}=\vec{x}_{org}+\delta$ over its legitimate counterpart $\vec{x}_{org} \in \mathbb{R}^{n\times m}$, where $n$ and $m$ denote the length of the signal and the number of channels, respectively, and $\delta$ is the adversarial perturbation.
Carlini and Wagner \cite{carlini2018audio} have demonstrated the effectiveness of this measure as a constraint in their optimization formulation for attacking a speech-to-text model (C\&W):
\begin{equation}
   \min \left | \delta \right |_{2}^{2}+\sum_{i} \vartheta_{i}.\mathcal{L}_{i}(\vec{x}_{org}+\delta_{i},\pi_{i}) \quad \mathrm{s.t.} \quad l_{\text{dB}_{\vec{x}}}(\delta) < \zeta 
   \label{eq:adv1}
\end{equation}
\noindent where $\pi_{i}$ refers to a character alignment (tokens without duplication) according to the target output phrase $\mathbf{y}_{i}$ in such a way that $\mathrm{Pr}(\pi_{i}|\mathbf{y}_{i})=\prod_{j}\mathbf{y}_{\pi^{j}}^{j}$. Additionally, $\mathcal{L}_{i}(\cdot)$ denotes the connectionist temporal classification loss \cite{graves2006connectionist}, and $\vartheta_{i}$ is a scaling factor. Finding an optimal value for $\zeta$ makes (\ref{eq:adv1}) brittle since it requires searching in an exponential space for a phrase $\mathbf{p}_{i}$, which should reduce to $\pi_{i}$ (after removing empty tokens). However, it has been shown that such a costly optimization formulation yields adversarial audios though sound very similar to $\vec{x}_{org}$, make the DeepSpeech system \cite{hannun2014deep} generate any target phrase pre-defined by the adversary \cite{carlini2018audio}. Since $\delta$ is not universal, slightly perturbing $\vec{x}_{adv}$ such as playback and recording over the air might override generating such a target phrase. In response to this issue, variants of expectation over transformation (EOT) have been developed as part of the optimization formulation inspired by \cite{athalye2018synthesizing}. Possible transformations are room impulse response, reverberation, and band-pass filters for truncating adversarial perturbation beyond human audible range \cite{yakura2018robust}. However, this strong approach is more costly than (\ref{eq:adv1}) and it fits well for short signals with a few corresponding phrases \cite{qin2019imperceptible}. The improved version of EOT has been recently introduced with a minor enhancement over the aforementioned distortion metric \cite{qin2019imperceptible}:
\begin{equation}
    10\log_{10}\left | \rho_{\delta} \right |^{2} - 10\log_{10}\left | \rho_{\vec{x}_{org}} \right |^{2}
\end{equation}
where $\rho$ denotes the power spectral density (PSD) function. They have also introduced a new formulation for the loss function according to the configuration of the Lingvo speech-to-text system \cite{shen2019lingvo} :
\begin{equation}
    \ell(\vec{x}_{i},\delta_{i},\mathbf{y}_{i})=\mathbb{E}_{t\sim \tau}\left [ \ell_{net}\left ( \mathbf{y_{i}}^{o},\mathbf{y_{i}}^{t} \right )+\alpha \ell(\vec{x}_{i},\delta_{i}) \right ]
    \label{eq:adv2}
\end{equation}
\noindent where $\alpha$ is a scalar and $\ell_{net}$ is the cross entropy loss which constrains over the normalized PSD function. Moreover, $\mathbf{y_{i}}^{o}$ and $\mathbf{y_{i}}^{t}$ denote the output and target phrases, respectively. This algorithm, which is known as robust attack, optimizes for the minimal $\delta_{i}$ over a set of $\tau$ transformations under varieties of room configurations. Similar minimization process has been implemented in a black-box scenario using a genetic algorithm (GA) \cite{taori2019targeted}. Specifically, this GA-based attack (GAA) incorporates a momentum mutation approach as well as gradient estimation in order to obtain optimal candidate populations associated with a predefined target phrase.     

\begin{figure*}[htpb!]
  \centering
  \includegraphics[width=0.75\textwidth]{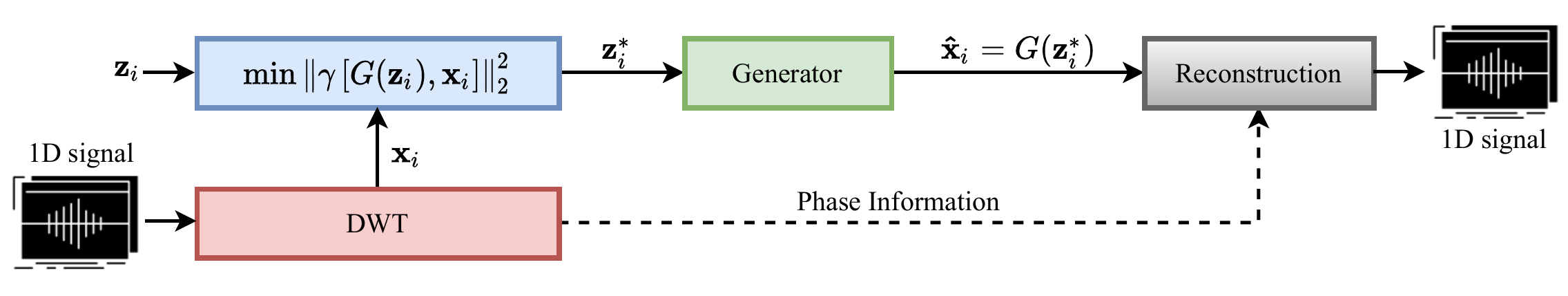}
  \caption{Overview of the proposed end-to-end defense-GAN approach. The 1D signal converted to a 2D-DWT spectrogram is denoted as $\mathbf{x}_{i}$ and the prior $p_{z}$ for $\mathbf{z}_{i} \in \mathbb{R}^{d_{z}}$ is $\mathcal{N}(0,0.4I)$. Additionally $\gamma\left [ \cdot \right ]$ is the chordal distance adjustment in the generalized Schur decomposition domain \cite{esmaeilpour2020detection} and $\hat{\mathbf{x}}_{i}$ represents the synthesized spectrogram from the generator. 1D signal is reconstructed using inverse DWT.} 
  \label{overview-defense}
  \vspace{-10pt}
\end{figure*}

While the fooling rate of the aforementioned adversarial attacks on DeepSpeech and Lingvo systems is almost 100\%, there are few studies on defense approaches for speech-to-text systems. This might be due to the immaturity of the end-to-end attack algorithms since several playbacks of the crafted adversarial signal over the air might bypass the achieved perturbations \cite{qin2019imperceptible}. Moreover, adversarial signals usually carry audible noises, even with $l_{\text{dB}_{\vec{x}}}({\delta})<0$, which makes their detection easier \cite{carlini2018audio}. However, reliable defense algorithms are still on demand against strong adversarial examples with less audible noises. Although there are some investigations for both proactive and reactive defense approaches \cite{zhang2019defending,das2018adagio}, they are characterized in a small scale.

In this paper, we propose a new reactive adversarial defense using a class-conditional generative adversarial network~\cite{mirza2014conditional}. We show that, our proposed defense scheme can be effective for large-scale systems such as DeepSpeech and Lingvo. The rest of the paper is organized as follows. In Section~\ref{sec:gfad} we provide a brief introduction to GANs focused on defense strategies for speech signals. Section~\ref{sec:pdgan} presents our defense approach that includes three major steps for removing potential adversarial perturbations from signals. Section~\ref{sec:exp} summarizes and discusses the experiments carried out on Mozilla common voice (MCV) and LibriSpeech datasets. Conclusion and perspective of future work are presented in the last section.

\section{GAN for Adversarial Defense}
\label{sec:gfad}
In a typical GAN configuration organized as a two-player minimax optimization problem \cite{goodfellow2014generative}, the generator network $G(\mathbf{z};\theta_{g})$ with $\mathbf{z} \in \mathbb{R}^{d_{z}}$ and training parameters $\theta_{g}$ learns to map from the designated distribution $p_{z}\sim \mathcal{N}(0,I)$ to $p_{g}$ as:
\begin{equation}
    \min_{G} \max_{D} \mathbb{E}_{\bold{x}\sim p_{r}(\bold{x})}\left [ \log D(\bold{x}) \right ]+\\
\mathbb{E}_{\bold{z}\sim p_{z}(\bold{z})}\left [ \log \left ( 1-D(G(\bold{z})) \right ) \right ]
    \label{gan1}
\end{equation}
\noindent where $p_{r}$ is the real sample distribution and $D(\mathbf{x}; \theta_{d})$ denotes the discriminator network with training parameters $\theta_{d}$. Upon carefully training $G(\mathbf{z};\theta_{g})$, it can generate seamless samples almost without recognizable perturbations compared to $\mathbf{x}_{i} \sim p_{r}$. In fact, the generator semantically learns real sample distribution and we should expect unnoticeable differences between the generated samples and random test inputs except for adversarial examples. Based on this idea, a reactive defense approach has been introduced by Samangouei et al.~\cite{samangouei2018defensegan}, which iteratively minimizes for $\left \| G(\mathbf{z})-\mathbf{x} \right \|_{2}^{2}$. Since the $L_{2}$ distance (or any other similarity metrics such as $L_{\infty}$) between crafted adversarial examples and their corresponding legitimate samples is fairly small, they extended their optimization problem subject to finding the most optimal $\mathbf{z}_{i}$. Unfortunately, this adversarial filtration defense scheme shatters gradient information and it can be easily disrupted by running a backward pass differentiable approximation (BPDA) attack \cite{athalye2018obfuscated}. On the contrary, the generator network can be trained to minimize the similarity between adversarial and legitimate samples where the discriminator iteratively learns to span possible adversarial manifolds \cite{lee2017generative}. Training such a defense-GAN requires exploring a massive adversarial subspace since not every attack algorithm generates a universal perturbation scale \cite{esmaeilpour2020detection}.

Autoencoder-based GAN (A-GAN) has also been investigated for defending speech emotion recognition models using long-short term memory networks \cite{latif2018adversarial}. This defense-GAN configuration introduces complex architecture for transforming a feature vector into another one aiming at bypassing potential adversarial perturbation. However, with the assumption of stable training without oversmoothing, this model might not necessarily enhance adversarial robustness against translation-invariant \cite{dong2019evading} or black-box attacks. However, these attacks are robust against low-level feature reconstruction using encoder-decoder blocks. In response to this issue and to the BPDA attack, we introduce a new defense-GAN architecture in a class-conditional framework which can be effectively used to increase the robustness of large-scale speech datasets and the state-of-the-art speech-to-text systems such as DeepSpeech and Lingvo.

\section{Proposed Defense Approach: CC-DGAN}
\label{sec:pdgan}
The proposed adversarial defense approach is made up of three steps, as shown in Fig.~\ref{overview-defense}: (i) generating signal representation; (ii) minimizing the relative chordal distance adjustment for the given input signal relative to $G(\mathbf{z}_{i})$; and (iii) signal reconstruction with the preserved phase information. We explain all these steps in detail as follows.

\subsection{2D Signal Representation}
\label{sec:signalRep}
Due to the high dimensionality of audio and speech signals, adversarial training either on single or multi-channel waveforms is very challenging and the model often undergoes complete collapse at early iterations. Therefore, a conventional approach in speech processing is to convert a given signal into a frequency-plot representation (spectrogram). Thus, as suggested by Esmaeilpour et al.~\cite{esmaeilpour2020unsupervised}, we divide the input signal into smaller chunks sampled at 16 kHz using discrete wavelet transform (DWT). Additionally, we set the frame length to 50 ms and use the complex Morlet mother function. Moreover, for enhancing the quality of the resulting spectrogram ($\mathbf{x}_{i}$ in Fig.~\ref{overview-defense}), we represent its magnitude in a logarithmic scale. It has been shown that these settings for spectrogram production outperform short-time Fourier transform both in terms of recognition accuracy and robustness against adversarial attacks \cite{esmaeilpour2019robust, DBLP:journals/corr/abs-2007-13703}. Since the dimensions of the generated $\mathbf{x}_{i}$ are not necessarily square, we bilinearly resize them to 128$\times$128 in compliance of computing the relative chordal distance in a non-Cartesian space.

\subsection{Chordal Distance Adjustment Minimization}
\begin{figure*}[htpb!]
  \centering
  \includegraphics[width=0.8\textwidth]{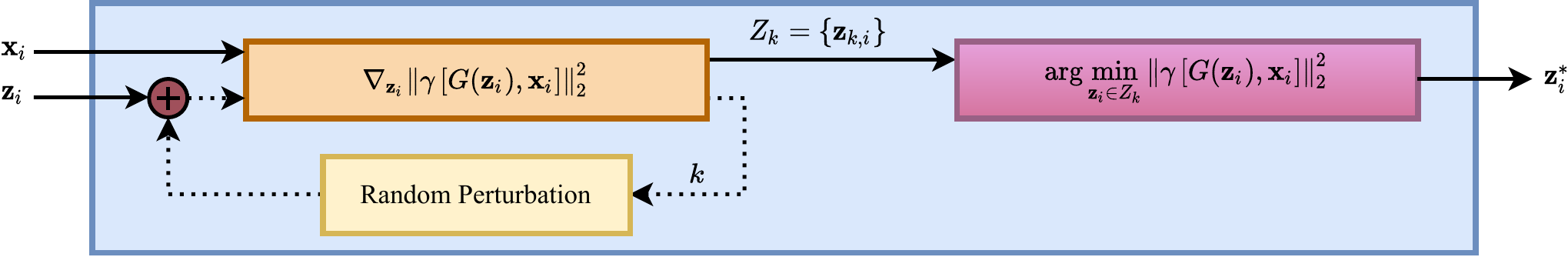}
  \caption{$k$ steps minimization for the chordal distance adjustment between $G(\mathbf{z}_{i})$ and $\mathbf{x}_{i}$. Similar to the predefined prior for $\mathbf{z}_{i}$, the random perturbation is also a function distributed over $\mathcal{N}(0,0.4I)$. The inner loop is shown in dotted line.}  
  \label{gammaMinimizer}
  \vspace{-10pt}
\end{figure*}
The chordal distance \cite{van1983matrix} is a metric that measures subspace adjacency for two similar samples in the domain of generalized Schur decomposition \cite{esmaeilpour2020detection}. This metric has been used for characterizing the existence of adversarial examples in subspaces different from legitimate and noisy samples \cite{esmaeilpour2020detection}. The chordal distance between an adversarial example $G(\mathbf{z}_{i})$ and $\mathbf{x}_i$ is:
\begin{equation}
    \mathrm{chord}\left ( \lambda\left [ G(\mathbf{z}_{i}) \right ], \lambda\left [ \mathbf{x}_{i} \right ] \right )=\frac{\left | \lambda\left [ G(\mathbf{z}_{i}) \right ] - \lambda\left [ \mathbf{x}_{i} \right ] \right |}{\sqrt{1+\lambda\left [ G(\mathbf{z}_{i}) \right ]^{2}}\sqrt{1+\lambda\left [ \mathbf{x}_{i} \right ]^{2}}} 
    \label{eq:chordalDistance}
\end{equation}
\noindent where $\lambda{\left [ \cdot \right ]}$ denotes the vector of eigenvalues for the designated spectrograms. Achieving a valid chordal distance between two spectrograms for ensuring their subspace adjacency in the generalized Schur decomposition enquires $\left \| G(\mathbf{z}_{i})-\mathbf{x}_i\right\|\simeq\xi_{i}$ where the threshold $\xi_{i}$ must be small according to the computed mean eigenvalue. For samples which lie in the same subspace, however with dissimilar spans, a minor translation is required in Eq.~\ref{eq:chordalDistance} to avoid ill-conditioned cases \cite{van1983matrix}. Specifically, for pencils $\vec{\mu}_{i}G(\mathbf{z}_{i})-\mathbf{x}_{i}$ and $\vec{\mu}_{i} \in \mathrm{diag}(\lambda\left [ G(\mathbf{z}_{i} \right ])/\mathrm{diag}(\lambda\left [ \mathbf{x}_{i} \right ])$, an adjustment $\gamma_{i} \left [ \cdot \right ]+\mathrm{chord}(\cdot)$ is needed in (\ref{eq:chordalDistance}), especially for samples with very small $L_{2}$ distance in Euclidean space \cite{esmaeilpour2020detection}.

Since the $\gamma_{i}\left[\cdot\right]$ adjustment is relatively large for an adversarial example $\mathbf{x}_{adv}$ \cite{esmaeilpour2020detection}, minimizing over $\left\|\gamma\left[G(\mathbf{z}_{i}) \right ],\gamma\left[\mathbf{x}_{adv}\right]\right\|_{2}^{2}$ projects $\mathbf{x}_{adv}$ onto the legitimate sample subspace distribution represented by $p_{g}$. However, we do not filter $\mathbf{x}_{adv}$, neither by conventional encoder-decoder blocks nor by low-level transformation operations. In fact, we find an optimal $\mathbf{z}_{i}^{*}$$\in$$\mathbb{R}^{d_{z}}$ through an iterative approach, then pass it to the generator for crafting a spectrogram very similar to the given $\mathbf{x}_{adv}$. This approach is depicted in Fig.~\ref{gammaMinimizer}, where the number of iterations for obtaining the optimal $\mathbf{z}_{i}^{*}$ is denoted by $k$. For avoiding possible ill-conditioned pencils \cite{van1983matrix}, we slightly perturb the candidate $\mathbf{z}_{k,i}$ with a random scalar and augment it with $\mathbf{z}_{i}$. Since $G(\mathbf{z};\theta_{g})$ is trained to support $p_{g}\approx p_{r}$, it considerably reduces the chance of generating spectrograms with adversarial perturbations. Therefore, the architectural design of both generator and discriminator has a crucial role. To this end, we propose simple yet effective class conditional architectures for reliable training.

\subsubsection{Class-Conditional Defense GAN (CC-DGAN)}
The proposed class-conditional defense GAN (CC-DGAN) is based on the vanilla GAN, where both the generator and the discriminator receive additional information on top of the noise vector $\mathbf{z}_{i}$ (i.e., $\mathbf{y}_{i}$) \cite{mirza2014conditional}. Unlike the baseline model (\ref{gan1}), the CC-DGAN requires class embeddings ($c$-embeddings) mainly for the generator network: $\log(1-D(G(\mathbf{z}|c=\mathbf{y})))$. This modification expands the learning space of the model at the risk of losing sample variety and mode collapse \cite{brock2018large}. However, we find that $c$-embeddings provide a considerable boost in computing character probability distribution at every frame of the given signal compared to regular GANs.

The proposed generator receives $\mathbf{z}_{i}$$\in\mathbb{R}^{128}$$\sim$$\mathcal{N}(0,I)$ in the first layer followed by a linear block with dimension $50+128$ and shared $c-\mathrm{embedding}=50$ \cite{PerezSVDC18} including $4\times 4 \times 16$ channels. There are two sequential residual blocks on top of the linear with $16 \rightarrow 4$ and $4 \rightarrow 1$ channels. The last hidden layer is a $128 \times 128$ non-local block with batch normalization and $\tanh$ activation function. The batch size is set $256$ with orthogonal initialization \cite{SaxeMG13}. Each residual block includes two linear ($128\times 128$) and three padded convolution ($3\times 3$ with stride 1) layers followed by upsampling, batch normalization, and ReLU activation function. In our discriminator network, the first layer requires RGB spectrogram $\mathbf{x}_{i}$$\in$$\mathbb{R}^{128\times 128\times 3}$. There is only one residual block in this network which contains two sequential $3 \times 3$ convolution layers with concatenation, ReLU, skip-$z$ \cite{brock2018large}, and average pooling. On top of the residual block, there is a $64\times 64$ non-local layer with 16 channels, ReLU, MaxPooling, and a linear logit layer ($\rightarrow 1$). Furthermore, we use both orthogonal regularization \cite{brock2017neural} and initialization \cite{SaxeMG13} for the entire weight vectors.

\subsection{Signal Reconstruction}
This is the third step of the proposed defense approach as shown in Fig.~\ref{overview-defense}. We reconstruct a given 1D signal with its own original phase information and the synthesized spectrogram $\hat{\mathbf{x}}_{i}$. Although synthesizing phase vectors with generative models is very challenging, there are some approaches for building them. However, they add audible hissing and whining noises to the signal. Signal reconstruction with original phase vectors often provides higher signal to noise ratio and this might help to more conveniently distinguish an adversarial example from a noisy signal \cite{koerich2020cross}. The reconstruction operation only requires running an inverse DWT with basic settings such as type of mother function, sampling rate, and frame length. We use the same settings mentioned in Section~\ref{sec:signalRep} with additional quantization filter for normalizing the achieved vectors. For simplicity, we assume that signals are single-channel.

\section{Experiments}
\label{sec:exp}
\begin{table*}[ht]
\footnotesize
\centering
\caption{Comparison of different defense approaches against white and black-box adversarial attacks for DeepSpeech and Lingvo victim models. Better results are shown in bold face. In the robust attack, $\Delta$ is the offset scalar: $\left \| \delta_{i} \right \|< \zeta_{i}+\Delta$ \cite{qin2019imperceptible} defined by the adversary.}
\begin{tabular}{|c||c|c|c|c|c|c|}
\hline
Model                                    & Attack                         & Defense     & Average $k$ & $\Delta$ & WER (\%)         & SLA (\%)         \\ \hline \hline
\multirow{6}{*}{DeepSpeech (Subset-MCV)} & \multirow{3}{*}{C\&W}          & A-GAN       & --          & --       & $23.98 \pm 2.14$ & $49.17 \pm 1.78$ \\ \cline{3-7} 
                                         &                                & Compression \cite{das2018adagio} & --          & --       & $17.41 \pm 3.07$ & $56.96 \pm 2.38$ \\ \cline{3-7} 
                                         &                                & Proposed CC-DGAN        & $67$        & --       & $\bold{05.37 \pm 2.66}$  & $\bold{78.15 \pm 1.08}$ \\ \cline{2-7} 
                                         & \multirow{3}{*}{GAA}           & A-GAN       & --          & --       & $18.54 \pm 5.31$ & $53.76 \pm 3.19$ \\ \cline{3-7} 
                                         &                                & Compression \cite{das2018adagio} & --          & --       & $\bold{03.81 \pm 1.16}$  & $\bold{70.14 \pm 5.72}$ \\ \cline{3-7} 
                                         &                                & Proposed CC-DGAN        & $54$        & --       & $03.97 \pm 0.44$  & $68.35 \pm 2.51$ \\ \hline \hline
\multirow{3}{*}{Lingvo (Subset-LS)}      & \multirow{3}{*}{Robust Attack} & A-GAN       & --          & $300$    & $21.23 \pm 4.79$ & $58.90 \pm 2.42$ \\ \cline{3-7} 
                                         &                                & Compression \cite{das2018adagio} & --          & $300$    & $19.34 \pm 3.91$ & $54.88 \pm 4.52$ \\ \cline{3-7} 
                                         &                                & Proposed CC-DGAN        & $59$        & $400$    & $\bold{07.26 \pm 3.08}$  & $\bold{67.36 \pm 1.77}$ \\ \hline
\end{tabular}
\label{table:comp}
\vspace{-10pt}
\end{table*}
We have evaluated the proposed defense (CC-DGAN) against three end-to-end adversarial attacks for both Mozilla's implementation of DeepSpeech \cite{MozillaImplementation} and Lingvo system \cite{shen2019lingvo}. These speech-to-text models are trained on Mozilla common voice (MCV) \cite{MozillaCommonVoiceDataset} and LibriSpeech \cite{panayotov2015librispeech} datasets, respectively. Both these benchmarking datasets contain above 1,000 hours of voice clips with various utterances. However, as a common practice \cite{carlini2018audio,qin2019imperceptible,taori2019targeted} we generate adversarial examples only for a portion of such datasets. We randomly select 11,500 and 6,000 samples from the MCV and LibriSpeech datasets for both training the CC-DGAN and running attacks, respectively. We organize these samples with their associated transcriptions into Subset-MCV and Subset-LS.   

We run white-box (C\&W) and black-box (GAA) adversarial attacks separately against the DeepSpeech model which uses rounds of long-short term memory blocks. We have randomly selected 1,000 samples from Subset-MCV with their original English transcriptions and we have targeted 10 different incorrect phrases (because these two attacks do not incorporate EOT) for effective attacking. Although these attacks directly optimize for achieving the minimum possible perturbation for the 1D signal, the DeepSpeech model first converts the given input into a Mel-frequency coefficient (MFC) representation. This adds more computational overhead to the attack algorithms and prohibits crafting adversarial examples for all the recordings in the dataset. The MFC layer splits the given speech signal into 50 frames per second which means the model can output up to 50 characters per second ($\mathbf{y}_{i}$). Therefore, this frame length is fairly enough for short signals with quite large transcripts. We extended these two attacks for targeting silence equivalent to generating empty tokens ($\epsilon$) for an additional 500 samples from the Subset-MCV. To this end, we updated the loss function as \cite{carlini2018audio}:
\begin{equation}
    \sum _{i}\max_{t \in \left \{ \epsilon \right \}}\left ( f(\vec{x})_{t}^{i}-\max_{\hat{t} \notin \left \{ \epsilon \right \}} f(\vec{x})_{\hat{t}}^{i}, 0 \right ),
    \quad f:\mathcal{X}^{50}\rightarrow [0,1]^{50\cdot \left | \pi \right |}
    \label{eq:loss2}
\end{equation}
\noindent where $50$ and $\mathcal{X}$ denote the number of frames and input space, respectively. Moreover, $\hat{t}$ is the target phrase defined by the adversary in replacement of the original transcript $t$. Targeting $\epsilon$ token is easier than lexical characters and considerably reduces the computational cost. For the Lingvo victim model using the robust attack, we also randomly select 1,000 samples from Subset-LS with their associated transcripts targeting one incorrect phrase (because it incorporates EOT) with the same settings as mentioned in \cite{taori2019targeted}. If the attack algorithm cannot exactly converge to a pre-defined target phrase, we replace it with another sample to keep the fooling rate at 100\%.

For evaluating the proposed CC-DGAN to counteract the three adversarial attacks, we firstly train them separately on Subset-MCV and Subset-LS. In order to avoid losing sample variety and to add bias to our generative models, we exclude those nominated samples for adversarial attacks. For both generator and discriminator networks, we use Adam optimizer \cite{kingma2014adam} with $\beta_{1}$=0, $\beta_{2}$=0.9, and a constant learning rate 2$\cdot$10$^{-4}$. We also run an exploratory search for finding the optimal number of steps required for $G(\mathbf{z};\theta_{g})$ over $D(\mathbf{x}; \theta_{d})$. We eventually opted to use two steps with decay rate 0.99 on two NVIDIA GTX-1080-Ti with 4$\times$11GB memory in addition to a 64-bit Intel Core-i7-7700 (3.6 GHz) CPU with 64GB of RAM.

As a common issue in adversarial training, the proposed CC-DGAN configuration also undergoes collapse at about 9.3k and 6.8k iterations for Subset-MCV and Subset-LS, respectively. For improving the stability of our models, we have employed spectral normalization \cite{miyato2018spectral} only for $G(\mathbf{z};\theta_{g})$. However, it turns out oversmoothing the generated spectrogram. For rectifying this issue, we replaced long speech signals with shorter recordings, randomly drawn from the original datasets. The final GAN models used for further evaluations are those achieved from the checkpoints prior to potential collapse, which happens at about 10k iterations on both subsets. The $k$-step optimization algorithm for achieving $\mathbf{z}_{i}^{*}$ is depicted in Fig.~\ref{gammaMinimizer} and finding a minimal value for it requires generalizable generative models. Regarding our experiments, for partially unstable and somewhat oversmoothed generators, $k$ never converges in less than 400 iterations.

For evaluating the performance of the proposed defense approach against the three aforementioned adversarial attacks, we use two metrics: (i) word error rate (WER), which is computed as $(I$+$S$+$D)/N\times$100 where $I$, $S$, $D$, and $N$ are the total number of insertions, substitutions, deletions, and reference words, respectively \cite{qin2019imperceptible}; (ii) sentence level accuracy (SLA), computed as $n_{c}/n_{tot}$ where $n_{c}$ is the number of samples which could achieve the correct transcript and $n_{c}$ is the total number of test speech signals. Table~\ref{table:comp} summarizes the results of our experiments, where both the SLA and WER are computed for the three defense algorithms. Specifically, these two metrics measure the performance of the defenses in producing phrases which reduce to correct transcriptions for the given adversarial signals. Note that, these two metrics while computed for the adversarial attacks, they measure fooling rates of the victim models in producing incorrect transcriptions as defined by the adversary. For consistent evaluation and in response to the raised concern of complete model vulnerability against end-to-end adversarial attacks \cite{carlini2018audio}, we set the SLA to 100\% for all defense algorithms. Since for effective evaluations we target 10 incorrect transcriptions for every speech signal under C\&W and GAA attacks, the reported results are averaged over 10 different runs. Table~\ref{table:comp} shows that for the majority of the cases, the proposed CC-DGAN outperforms both simple compression and complex autoencoder-based GAN (A-GAN) in removing potential adversarial perturbations from speech signals and achieving lower WER and higher SLA. The only exception is for the GAA attack, which implements approximated gradient estimation, where simple compression achieves a slightly better performance than the proposed CC-DGAN. We noticed that for such attack, doubling $k$, reduces the WER in about 1.09$\%$ and increases the SLA in around 2.58$\%$ compared to $k$=54. For better investigating this issue, we attacked both victim models with the BPDA attack and measured the performance achieved by the proposed defense GAN. Our investigation on the same crafted adversarial examples uncovered the effectiveness of this attack on the CC-DGAN. More specifically, for reaching almost the same WER and SLA reported in Table~\ref{table:comp}, $k$ should be increased 2.37 and 3.12 times more for DeepSpeech and Lingvo systems, respectively.

\section{Conclusion}
In this paper, we proposed a new defense algorithm for securing advanced DeepSpeech and Lingvo systems against three end-to-end white-box and black-box adversarial attacks. The proposed CC-DGAN uses simple architectures for both the generator and discriminator with few residual blocks and a reconstructor module. This module regenerates a test input speech with the synthesized DWT spectrogram and its original phase information for seamless reconstruction. The experimental results on subsets of MCV and LibriSpeech datasets have shown that, the proposed defense approach considerably outperforms other defense algorithms for the majority of the cases in terms of achieving lower WER and higher SLA. Since the performance of our defense approach is highly dependent on the generalizability of the CC-DGAN, we are inclined to improve its stability and increase its generalizability in our future studies.


\clearpage
\balance
\bibliographystyle{IEEEbib}
\bibliography{refs}
\vfill

\end{document}